\documentclass[letter,traditabstract]{aa}
\usepackage{graphicx}
\usepackage{txfonts}
\usepackage{natbib}
\usepackage{enumitem}
\bibpunct{(}{)}{;}{a}{}{,}

 \newcommand{\mic}{$\mu$m}
 \newcommand{\mics}{$\mu$m~}

\begin{document}

\title{The dust content of high-z submillimeter galaxies revealed by \textit{Herschel} \thanks{\textit{Herschel}
is an ESA space observatory with science instruments provided by European-led Principal 
Investigator consortia and with important participation from NASA.
}}

\author{
P. Santini\inst{1}
\and
R. Maiolino\inst{1}
\and
B. Magnelli\inst{2}
\and
L. Silva\inst{3}
\and
A. Grazian\inst{1}
\and
B. Altieri\inst{4}
\and
P. Andreani\inst{3,5}
\and
H. Aussel\inst{6}
\and
S. Berta\inst{2}
\and
A. Bongiovanni\inst{7,8}
\and
D. Brisbin\inst{9}
\and
F. Calura\inst{3,10}
\and
A. Cava\inst{7,8}
\and
J. Cepa\inst{7,8}
\and
A. Cimatti\inst{11}
\and
E. Daddi\inst{6}
\and
H. Dannerbauer\inst{6}
\and
H. Dominguez-Sanchez\inst{12}
\and
D. Elbaz\inst{6}
\and
A. Fontana\inst{1}
\and
N. F{\"o}rster Schreiber\inst{2}
\and
R. Genzel\inst{2}
\and
G. L. Granato\inst{3}
\and
C. Gruppioni\inst{12}
\and
D. Lutz\inst{2}
\and
G. Magdis\inst{6}
\and
M. Magliocchetti\inst{13}
\and
F. Matteucci\inst{14}
\and
R. Nordon\inst{2}
\and
I. P{\'e}rez Garcia\inst{4}
\and
A. Poglitsch\inst{2}
\and
P. Popesso\inst{2}
\and
F. Pozzi\inst{11}
\and
L. Riguccini\inst{6}
\and
G. Rodighiero\inst{15}
\and
A. Saintonge\inst{2}
\and
M. Sanchez-Portal\inst{2}
\and
L. Shao\inst{2}
\and
E. Sturm\inst{2}
\and
L. Tacconi\inst{2}
\and
I. Valtchanov\inst{2}
}

\institute{\centering \vskip -10pt \small \it (See online Appendix \ref{sect:affiliations} for author affiliations) }

   \offprints{P. Santini, \email{paola.santini@oa-roma.inaf.it}}

   \date{Received .... ; accepted ....}
   \titlerunning{The dust content in high-z SMGs}

   \abstract{We use deep observations taken with the Photodetector Array Camera and Spectrometer (PACS), on board the \textit{Herschel} satellite 
   as part of the PACS evolutionary probe (PEP) guaranteed project along with submm ground-based observations to measure
   the dust mass of a sample of high-z submillimeter galaxies (SMGs).
	We investigate their dust content relative to their stellar and gas masses, and compare them with local star-forming galaxies.
	High-z SMGs are dust rich, i.e. they have higher dust-to-stellar mass ratios  compared to local spiral galaxies (by
	a factor of 30) and also compared to local ultraluminous infrared galaxies (ULIRGs, by a factor of 6).
	This  indicates that the large masses of gas typically hosted in SMGs have already
	been highly enriched with metals and dust. Indeed, for those SMGs whose gas mass is measured, we infer  dust-to-gas ratios similar
	or higher than local spirals and ULIRGs. However, similarly to other strongly star-forming galaxies in the local Universe and at high-z, 
	SMGs are characterized by gas metalicities {\it lower} (by a factor of a few)
	than local spirals, as inferred from their optical
	nebular lines, which are generally ascribed to infall of metal-poor gas.
	This is in contrast with the large dust content inferred from the far-IR and submm data. In short,
	the metalicity inferred from the dust mass is much higher (by more than an order of magnitude) than that inferred
	from the optical nebular lines. We discuss the possible explanations of this discrepancy and the possible implications
	for the investigation of the metalicity evolution at high-z.}


\maketitle

\section{Introduction}

Understanding the evolution of the dust properties and the dust content of galaxies through the cosmic epochs is  crucial 
for constraining galaxy evolutionary scenarios. The amount of dust has important implications for the star formation (SF). Dust allows the formation of low-mass stars in low-metalicity environments while it inhibits the formation of massive stars, hence affects the IMF \citep[e.g.][]{omukai05}.
Dust greatly enhances the formation of many molecules whose transitions provide the main cooling mechanism for molecular clouds to 
form stars. Hence  dust is one of the pre-requisites for enhanced SF. 
Dust also affects  the detectability of high-z galaxies:  dust extinction reduces the detectability of high-z galaxies
in the rest-frame UV-optical bands, while dust thermal emission allows the detection of high-z galaxies at mm/far-IR wavelengths.
The dust content can also be used as a proxy of the metalicity of the ISM, because refractory elements in the ISM
are mostly depleted into dust grains.

The ESA \textit{Herschel} Space Observatory \citep{pilbratt10} offers the possibility to investigate the dust emission in high-z star-forming
galaxies and in particular to measure their dust content by modeling the far-IR to submm spectral energy distribution (SED). We here present a first investigation of dust mass evolution  by focusing on a sample of high-z submillimeter galaxies (SMGs). 
They are very luminous ($\sim 10^{13} L_\odot$) \citep[e.g.,][]{hughes98,pope06}, high redshift ($z \sim$ 1$-$3.5) \citep{chapman05,pope06}, gas rich and compact, massive galaxies \citep{greve05,tacconi08}. 
Their star formation rates (SFR) are exceptionally high ($\sim 10^3 M_\odot yr^{-1}$) and are thought to contribute significantly to the cosmic SF at $z \sim$ 2$-$3 \citep{chapman05}. Thanks to their high luminosities they provide an excellent laboratory to investigate the evolution of dust in high-z
galaxies. 
To measure their dust content we use data taken with PACS \citep{poglitsch10}
as part of the PACS Evolutionary Probe  
guaranteed time key program combined with ground-based mm and submm data. By comparing the inferred dust masses with other properties of the SMGs,
and with local galaxies, we infer  important information on the physics of these high-z systems. 
We adopt the $\Lambda$-CDM concordance cosmological model (H$_0$ = 70 km/s/Mpc, $\Omega_{\small M} $ = 0.3 and $\Omega_{\Lambda}$ = 0.7).

\section{Sample and data analysis}

We used deep 100 and 160 \mics PACS GTO observations of the field GOODS-N 
and of the lensing cluster Abell2218.
Observations, data reduction and source extraction are described in detail 
in \cite{berta10}. 
As mentioned above, we used a subset of PACS sources with
submm detections taken 
from the literature. 
The 
SMG sample and counterparts association are described in \cite{magnelli10}. 
The sample is restricted to sources with
secure redshifts (all spectroscopic, except for three sources with good photo-z) 
derived from robust multi-wavelength associations. 
A few SMGs in the GOODS-N field have two optical, and eventually PACS, 
counterparts at the same redshift. For these  we considered the total system of the two galaxies, which are likely interacting \citep{pope06},
by summing their IR fluxes and  stellar masses.
We  excluded three galaxies (GN19, GN19b, GN30) 
with only two photometric points at 
$\lambda$ $>$100\mics
because their SED could not be properly constrained.
Finally, we excluded a galaxy (azt23) 
with a peculiar SED,
suspected to be dominated by a powerful AGN even at far-IR wavelengths. 
We ended up with 12 SMGs in GOODS-N and 5 
in Abell2218, three of the latter are different images of the same lensed galaxy. 
All our galaxies have PACS (100 and 160\mic) and SCUBA 850\mics photometry, and nine also have SCUBA 450\mics and/or AzTEC 1.1mm and/or MAMBO 1.2mm detections. 
Redshifts range from 0.5 to 4, with a median value of 2. 

Both to expand the sample and to check that our dust masses are not subject to systematics associated with the specific
set of data, we also included six SMG galaxies
observed by \cite{kovacs06} at 350$\mu$m with SharcII, which have at least two additional (sub)mm photometric points, and one MIPS  (both 70 and 160\mic) detected SMG from the \cite{yan10} sample, with also a MAMBO 1.2mm observation.

Dust masses were derived by fitting and normalizing the 100\mic -to-(sub)mm photometry to the SED library generated by GRASIL \citep{silva98}. This is a chemospectrophotometric code that uses realistic and physically grounded stellar and dust distributions and geometries and performs the full radiative transfer calculations,
eventually providing the complete SED of various galaxy models. Recently,
GRASIL has been convolved with models of dust evolution in galaxies \citep{calura08,schurer09}. 
The model carefully takes into account the different emissivities of different populations of dust grains, which also depend on their size,
composition, and temperature. This effectively results in an average emissivity index $\beta \approx 2$, consistent with various
previous studies \citep{silva98,draineli07,clements10}. 
The dust masses inferred by us   broadly agree with those obtained by \cite{michalowski10a}, who also use GRASIL, although
the lack of data at $\rm 24\mu m <\lambda < 850\mu m$ in most objects prevents them from tightly constraining the far-IR bump.

The inferred dust mass is compared with the stellar mass, gas mass, and gas metalicity as inferred for a subset of the objects in
our total sample by  ancillary data presented in the Appendix. We also compared the dust properties of SMGs with those of local
star-forming galaxies,
both spirals and ULIRGs, whose properties are reported in the Appendix. Here we only mention that the  methods to derive the dust mass, stellar mass, gas mass, and metalicity are  the same for all samples, so that their comparison is done in a fully consistent way.

\begin{figure}[!t]
\resizebox{\hsize}{!}{\includegraphics[angle=0]{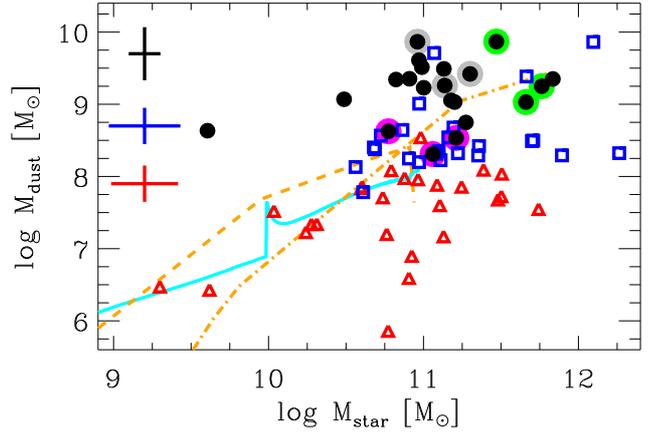}}
\caption{Stellar mass versus dust mass. 
Blue squares show local ULIRGs, red triangles refer to local spirals, while black circles correspond to high-z SMGs. 
Large gray, magenta and green circles mark respectively photo-z, 
the triple image in Abell2218 and blended SMG systems. 
Median $1\sigma$ error bars for the different samples (same color code) are shown on the left.   
The solid cyan and dashed (dot-dashed) orange lines show the 
predictions of \cite{calura08} model 
for spirals and proto-ellipticals with mass of $\rm 10^{11}$ ($10^{12}$) $\rm M_{\odot}$.
}
\label{fig:mdustmstar}
\end{figure}

\section{Results and discussion}

\begin{figure}[!t]
\resizebox{\hsize}{!}{\includegraphics[angle=0]{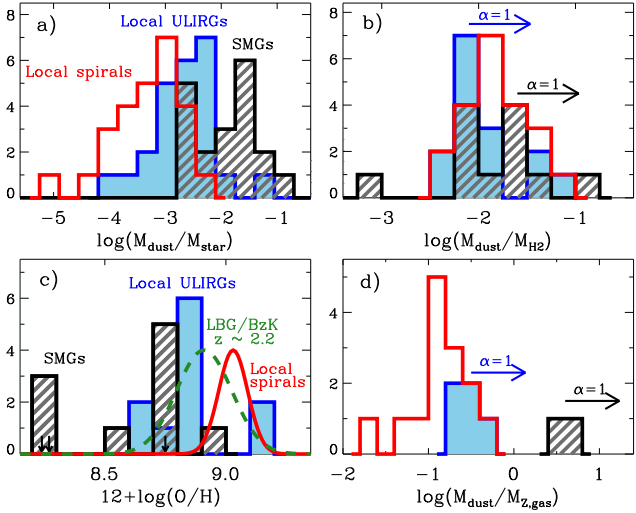}}

\caption{\textit{Panels a), b), d)}: $\rm M_{dust}/M_{star}$, $\rm M_{dust}/M_{H_2}$ and $\rm M_{dust}/M_{Z,gas}$ distributions, respectively, for
high-z SMGs (black hatched),  local spirals (SINGS, red open) and local ULIRGs (blue shaded). 
$\rm M_{H_2}$ is computed adopting $\rm \alpha = 4.3$. Adopting $\rm \alpha = 1$ moves the black and blue histograms
by the amount indicated by the arrows. 
\textit{Panel c)}: metal abundances for the sample of high-z SMGs of \cite{swinbank04} (black hatched
 histogram) and that of local ULIRGs of \cite{rupke08} (blue shaded
 histogram). 
Submillimeter galaxies  in common with our sample are marked by arrows. The red solid and green dashed lines
show the metalicity distribution for local spirals (SDSS) and $z \sim 2.2$ star-forming galaxies \citep{mannucci10}, respectively,
with stellar masses similar to those of our SMGs.  
}
\label{fig:histo}
\end{figure}

Figure~\ref{fig:mdustmstar} shows the stellar mass vs the dust mass for high-z SMGs (black circles), local ULIRGs
(blue squares) and local spirals (red triangles).  The dust masses estimated for local spirals nicely agree with those
obtained through the very detailed analysis of \cite{draine07} of the same sample. Most of them also agree fairly well with the evolutionary pattern for spiral galaxies predicted by the model of \cite{calura08} (cyan solid line).  
The dust masses were compared with those obtained by the more common assumption of a single temperature dust distribution. Assuming an emissivity index $\beta = 2$ (for consistency with GRASIL) and a single, absolute average
emissivity of dust at 125$\mu$m of 2.64 
\citep{dunne03} or 1.87m$^2$ kg$^{-1}$ \citep{lidraine01}, dust masses are on average 0.3 or 0.15 dex, respectively, lower than those inferred through GRASIL \citep[100\mics flux not used in the fitting as in][]{magnelli10}. 
Independently of the adopted method, there are clear, systematic differences in terms of dust content for the different samples.  The orange dashed and dot-dashed
lines in Fig.~\ref{fig:mdustmstar} show the expected evolution of proto-ellipticals \citep{calura08} with different mass. 
These models predict  a larger dust content during the active phase
of elliptical formation compared to normal spirals. Nonetheless, they can hardly reproduce the large $\rm M_{dust}/M_{star}$
observed in most SMGs. 
The dust-to-star excess in SMGs emerges more clearly in Fig.~\ref{fig:histo}a, which shows the distribution of $\rm M_{dust}/M_{star}$  for the different samples. Local ULIRGs
have a larger dust content than spirals for a given stellar mass, with  $\rm M_{dust}/M_{star}$  a factor of $\sim$4 higher on average. This result is in contrast with \cite{clements10}, who claimed that the dust content in ULIRGs can be simply explained with the combination of two spirals (the different result  probably arises because Clements et al. use only K-band photometry as a proxy of the stellar mass, while we perform the full SED fitting). Submillimeter galaxies are much more extreme: their $\rm M_{dust}/M_{star}$  is on average higher by a factor of $\sim$30 than in local spirals, and by a factor of $\sim$6  than in local ULIRGs. This large dust content is difficult to account for with the galaxy evolutionary model of \cite{calura08}.

High-z SMGs are known to be very gas rich, with gas fractions approaching 50\% \citep{tacconi08}.
The high $\rm M_{dust}/M_{star}$ values 
suggest that the large gas masses hosted by SMGs have already been highly enriched
with metals and dust; i.e. the excess of dust mass in SMGs is not  
a consequence of SF episodes following recent inflows of metal-poor gas accreted during mergers \citep[in contrast with the expectation of some models, ][]{montuori10}.
This can be tested by measuring the dust-to-gas 
ratio for those SMGs with CO millimetric
data. 
By assuming the same CO-to-H$_2$ conversion
factor $\rm \alpha = 4.3~M_\odot (K\ km\ s^{-1}\ pc^2)^{-1}$ for all samples (typically applied to normal spirals) the resulting distribution of
$\rm M_{dust}/M_{H_2}$ is shown in Fig.~\ref{fig:histo}b, similar for the three samples. If one adopts for ULIRGs and SMGs the
conversion factor $\rm \alpha \sim 1~M_\odot (K\ km\ s^{-1}\ pc^2)^{-1}$ \citep[thought to be more appropriate for these classes of objects,
][]{solomon05,tacconi08}, then ULIRGs and SMGs have $\rm M_{dust}/M_{H_2}$ significantly larger than local spirals, as indicated
by the arrows. It is not possible to infer the total gas mass for SMGs by including the fraction of HI, 
because this is not observed; but similarly to local ULIRGs, the bulk of the gas is probably in the molecular phase
\citep{sanders96} in contrast to spirals, which may have a considerable fraction of gas in the atomic phase.
As a consequence, by including $\rm M_{HI}$ to obtain the dust-to-total gas ratio, the resulting $\rm M_{dust}/M_{gas}$ distribution of 
ULIRGs and SMGs would be further skewed towards higher values relative to spirals.
A dust-to-gas ratio higher than in spirals has also
been observed in the central (dense) region of local luminous infrared galaxies (LIRGs) \citep{wilson08}. It can  possibly be ascribed to enhanced dust production 
by SNe and by the first generation of AGB stars in young stellar systems \citep[e.g.][]{michalowski10},
and/or to higher accretion of metals on dust grains in
these dense environments. In either cases,  the finding that the dust-to-gas ratio in SMGs is {\it not} lower than in local spirals supports the interpretation of their ``dust richness'' (relative to the stellar mass) in terms of ``gas richness'',
possibly along with a higher $\rm M_{dust}/M_{H_2}$.
Summarizing, the high dust content observed
in SMGs mirrors their high gas content, the latter with a dust-to-gas ratio similar or higher than normal spiral
galaxies.

The latter result is  in contrast with the low gas metalicity observed in SMGs though. Indeed, based on their optical nebular lines 
SMGs have metalicities similar or lower than other strongly star-forming systems, 
such as ULIRGs in the local Universe \citep{rupke08},
as well as Lyman break galaxies (LBGs), BzKs \citep[see][]{daddi04} and ULIRGs at high redshift \citep{erb06,mannucci10,caputi08}, which in turn are lower
than local, moderately star-forming spirals with the same stellar mass. This is illustrated in Fig.~\ref{fig:histo}c, where the
metalicity distribution of different classes of objects was measured with the same method (based on optical nebular lines) and adopting the same metalicity
scale as discussed in the Appendix, and within similar stellar mass interval ($\rm M_{\star}\sim 10^{11}M_{\odot}$).
For the SMGs we exploited the sample with rest-frame optical spectroscopy given in
\cite{swinbank04}, which unfortunately overlaps with our sample only for three objects (marked with black arrows). That galaxies with enhanced star formation are characterized by lower metalicity has been noted by various
authors \citep{kewley08, rupke08, caputi08}. 
Recently \cite{mannucci10} showed that the anticorrelation between metalicity and SFR actually accounts for most of the evolution of the mass-metalicity relation at high redshift: high-z galaxies (at least up to \textit{z}$\sim$2.5) appear more metal-poor simply because characterized by higher SFRs as a consequence of selection effects. 
This anticorrelation between SFR and metalicity is ascribed to inflow of metal-poor gas that boosts SF in galaxies. 
However, the inferred ``low'' gas metalicities in SMGs are in striking contrast with the  high dust content inferred from their far-IR emission.

\begin{figure}[!t]
\resizebox{\hsize}{!}{\includegraphics[angle=0]{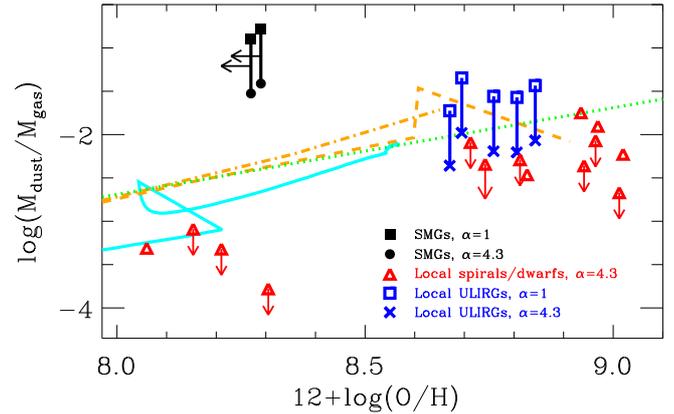}}
\caption{
Dust-to-gas ratio versus metalicity. The black and blue symbols show high-z SMGs and local ULIRGs, respectively, by assuming  different  conversion factors  as indicated by the legend.  
Red triangles refer to local spirals.  
The green dotted line shows the trend expected by Eq.~\ref{eq:draine}. 
The solid cyan and dashed (dot-dashed) orange lines show the 
evolutionary tracks predicted by the model of  \cite{calura08} 
for spirals and proto-ellipticals with mass of $\rm 10^{11}$ ($10^{12}$) $\rm M_{\odot}$.}
\label{fig:metal}
\end{figure}

In Fig.~\ref{fig:metal} we further investigate this discrepancy by comparing the dust-to-gas ratio with the gas metalicity.
Red triangles are all local galaxies for which the complete required information is available;
most of the upper limits are due to the lack of detection or of data
for the molecular gas. The dotted line shows the trend expected by assuming that the dust content scales linearly with
metalicity following the relation \citep{draine07}
\begin{equation}
\rm M_{dust}/M_{gas}\approx 0.01\times(O/H)/(O/H)_{MW}. 
\label{eq:draine}
\end{equation}
Actually, \cite{draine07} and \cite{hunt05} have shown (on a wider sample) that the relation traced by real galaxies
is much steeper than expected by Eq.~\ref{eq:draine}, as also hinted by the (much smaller) subset of local spirals/dwarfs
shown in Fig.~\ref{fig:metal}. Local ULIRGs (blue symbols) tend to have higher dust-to-gas ratios for a given metalicity,
the deviation depending on whether the assumed CO-to-H$_2$ conversion factor is $\alpha=1$ (open squares) or $\alpha=4.3$ (crosses). In the case of SMGs (black symbols) there are only two objects for which the complete required information is available (whose gas metalicity is actually an upper limit),  shown both in the case of  $\alpha=1$ (solid squares) and $\alpha=4.3$ (solid circles). However, it is clear that their dust content is far higher than expected from their metalicity.  The HI mass is not included for ULIRGs and SMGs, but the effect is very small \citep[0.1 dex, ][]{sanders96}.  The orange and cyan tracks in Fig.~\ref{fig:metal} show the expected evolution of the dust content and metalicity as predicted by \cite{calura08} for ellipticals and spirals, where it is evident 
that the large discrepancy observed in SMGs cannot be accounted for by models. 
In short, in SMGs the mass of metals inferred from the dust content is much larger
than the mass of metals inferred from  the gas phase.

To further quantify this discrepancy, in Fig.~\ref{fig:histo}d we show the distribution of $\rm M_{dust}/M_{Z,gas}$,
where $\rm M_{Z,gas}$ is the mass of metals in the gas phase (i.e. $\rm Z\times M_{gas}$). We  assumed for
all objects a conversion factor $\alpha=4.3$, the 
arrows showing the effect of using $\alpha=1$ for 
ULIRGs and SMGs.
Clearly in SMGs $\rm M_{dust}/M_{Z,gas}$ is over one order of magnitude higher than in spirals and well above unity.
Local ULIRGs share a similar problem, but not as extreme as SMGs. 
When computing $\rm M_{Z,gas}$, we assume the same metal abundance for the neutral  and  ionized  gas phases. This can be false, as
the former may be characterized by  lower \citep[by a factor of $\sim$7, ][]{leboutillier09} abundances than the latter. However, this effect
does not change our results. Indeed, dwarf low-metalicity
galaxies (see the Appendix), where most of the gas is in the atomic phase, would move towards higher values of  $\rm M_{dust}/M_{Z,gas}$ by
0.85 dex, while normal spiral galaxies, whose atomic gas fraction is around 50\%, would only move by 0.24 dex. The ratio  $\rm
M_{dust}/M_{Z,gas}$ in spiral galaxies and dwarfs would
still be lower than in  SMGs, where the bulk of the gas is expected to be in the
molecular phase (as discussed above). 

There are some possible scenarios that could explain this puzzling result.
One possibility is that dust masses are erroneously estimated because the dust properties in SMGs are very different from those assumed in our models. The average dust emissivity may be higher. 
However, this difference would have to be quite dramatic, because the dust masses are off by more than an order of magnitude with respect to those expected by the metalicities. In ULIRGs (extreme starbursts) the
emissivity inferred by \cite{clements10} is similar to that assumed here.

Another possibility is that as mentioned above the high density in SMGs favors the growth of dust out of metals in the ISM, which also decreases the effective metalicity in the gas phase. 
However, the standard dust depletion factors in the diffuse ISM are already high, with most metals already locked into dust.
Hence, for a given metalicity the dust mass cannot grow by more than a factor of about two through this mechanism.

Finally, the disagreement between the dust content and the gas metalicity can be caused by dust obscuration.
Likely, the bulk of the gas in SMGs is metal rich.
The associated dust richness and
the compact configuration make most of the ISM optically thick at visual wavelengths.
As a consequence, the optical nebular lines used to infer the gas metalicity only probe the outer parts of the star-forming regions, which are probably 
more metal-poor.
Paradoxically, the higher metalicity of these dense systems and their associated high dust content may produce the apparent effect of  metal-poorness when observed at optical wavelengths. 
This scenario is supported by the findings of \cite{santini09}, who showed that luminous IR galaxies have higher mid-IR-based SFR estimates compared to those obtained by correcting the optical-UV light.

If confirmed, the latter two scenarios would not support the current interpretation of the low metalicity observed
in ULIRGs and SMGs in terms of inflow of metal-poor gas boosting the star formation \citep[e.g.][]{montuori10}. More generally,
if the discrepancy between gas and dust metal mass also applies  to other populations of high-z starburst galaxies (e.g.
BzKs and LBGs) this would prompt  a general revision of the interpretation of the apparently low metalicities observed
in high-z systems.

\begin{acknowledgements}
PACS has been developed by a consortium of institutes led by MPE (Germany) and including UVIE
(Austria); KU Leuven, CSL, IMEC (Belgium); CEA, LAM (France); MPIA (Germany); INAF-IFSI/
OAA/OAP/OAT, LENS, SISSA (Italy); IAC (Spain). This development has been supported by the
funding agencies BMVIT (Austria), ESA-PRODEX (Belgium), CEA/CNES (France), DLR (Germany),
ASI/INAF (Italy), and CICYT/MCYT (Spain).
This work was supported by ASI through grant I/005/07/0.
\end{acknowledgements}

\bibliographystyle{aa}

\Online

\begin{appendix}

\section{Authors affiliations}\label{sect:affiliations}

\begin{enumerate}[label=$^{\arabic{*}}$]
\item INAF - Osservatorio Astronomico di Roma, via di Frascati 33, 00040 Monte Porzio Catone, Italy.
\item Max-Planck-Institut f\"{u}r Extraterrestrische Physik (MPE), Postfach 1312, 85741 Garching, Germany.
\item INAF-Osservatorio Astronomico di Trieste, via Tiepolo 11, 34131 Trieste, Italy.
\item Herschel Science Centre; European Space Astronomy Centre.
\item ESO, Karl-Schwarzschild-Str. 2, D-85748 Garching, Germany.
\item Laboratoire AIM, CEA/DSM-CNRS-Universit{\'e} Paris Diderot, IRFU/Service d'Astrophysique, B\^at.709, CEA-Saclay, 91191 Gif-sur-Yvette Cedex, France.
\item Instituto de Astrof{\'i}sica de Canarias, 38205 La Laguna, Spain.
\item Departamento de Astrof{\'i}sica, Universidad de La Laguna, Spain.
\item Department of Astronomy, 610 Space Sciences Building, Cornell University, Ithaca, NY 14853, USA.
\item Jeremiah Horrocks Institute for Astrophysics and Supercomputing, University of Central Lancashire, Preston PR1 2HE, UK. 
\item Dipartimento di Astronomia, Universit{\`a} di Bologna, Via Ranzani 1, 40127 Bologna, Italy.
\item INAF-Osservatorio Astronomico di Bologna, via Ranzani 1, 40127 Bologna, Italy.
\item INAF-IFSI, Via Fosso del Cavaliere 100, 00133 Roma, Italy.
\item Dipartimento di Astronomia, Universit{\`a} di Trieste,
via G. B. Tiepolo 11, 34143 Trieste, Italy.
\item Dipartimento di Astronomia, Universit{\`a} di Padova, Vicolo dell'Osservatorio 3, 35122 Padova, Italy.\end{enumerate}

\section{Ancillary data} \label{ancillary}

Here we first describe the ancillary data used for SMGs (stellar mass, gas mass and metalicity) and afterwards the local
samples used for comparison.

\subsection{Submillimeter galaxies}

Stellar masses of SMGs
are computed by fitting the optical-to-near-IR photometry compiled by the PEP Team to synthetic models, which is fully described in
\cite{fontana06}, fixing the redshifts to those of the SMGs.
For consistency with the theoretical models of \cite{calura08} with which we compare our
results, we adopt a Scalo IMF, which implies stellar masses higher than those calculated assuming a Salpeter IMF by a factor 1.15. 

When possible, we  obtained information on the gas mass. 
For the SMG sample we were able to find CO millimetric data for 12 galaxies (the three measurements of the triple image
of the lensed galaxy in Abell2218 are counted once) \citep{bothwell09,kneib05,daddi09,greve05,knudsen09}. 
To derive the molecular gas masses we followed the prescriptions given in the associated references,
but we normalized them to the same CO-to-$\rm H_2$ conversion factor. In particular we show the results assuming (conservatively)
a CO-to-H$_2$ conversion factor  $\rm \alpha = 4.3~M_\odot (K\ km\ s^{-1}\ pc^2)^{-1}$, which applies to normal spirals; but we also
show the effect of assuming $\rm \alpha = 1~M_\odot (K\ km\ s^{-1}\ pc^2)^{-1}$, which is thought to be more appropriate for ULIRGs
and SMGs \citep{solomon97,tacconi08}.

We inferred the gas metalicities through the optical nebular line emission by using the strong-line methods. Within this context, and
especially when comparing different samples, it
is most important to use consistent (inter-)calibrations and, if possible, use the same diagnostic emission line ratios.
Here we adopted the calibrations obtained by \cite{nagao06} and refined by \cite{maiolino08}, who inter-calibrate all strong line diagnostics
to the same scale. For the SMGs we used as metalicity diagnostic the ratio
[NII]/H$\alpha$ given in \cite{swinbank04}. An AGN may easily affect the nebular line ratios and make them deviate from the metalicity calibration for star-forming galaxies. For this reason we removed targets classified as AGNs in \cite{swinbank04}.
We also removed one object with [NII]/H$\alpha$ $=$0.65, because such a high ratio is only observed in AGNs or shocked regions. In total, we obtain gas metalicities for a sample of 10 SMGs, three of which also belong to our SMG sample with dust masses. The [NII]/H$\alpha$ ratio of these objects may still be affected by an unrecognized AGN, but this would make our results conservative, because the contribution by an unrecognized AGN would make the gas metalicity apparently higher than its actual value.

\subsection{Local galaxies}

To compare the dust content in high redshift SMGs with normal and actively star-forming galaxies in the local Universe, we assembled two sets of data:

\begin{itemize}
\item {a sample of 26 local spirals out of the SINGS sample \citep{kennicutt03,dale07,draine07}
with full multi-wavelength photometry and  also with submm data for a proper dust mass measurement;}
\item {a sample of 24 local ULIRGs taken from the \cite{clements10} sample with at least three submm photometric points.
Optical and near-IR fluxes needed to compute the stellar mass of these galaxies where collected from public archives. 
}
\end{itemize}

Stellar masses were inferred by  the same method and IMF as SMGs.

For the local spirals, we took the $\rm HI$ and $\rm H_2$ mass estimates from  \cite{kennicutt03},
while for the ULIRGs we collected $\rm H_2$ mass estimates from \cite{sanders88}, \cite{sanders91}, \cite{solomon97}, \cite{evans02} and \cite{papadopoulos10}. 
Here we also assumed  a CO-to-H$_2$ conversion factor  $\rm \alpha = 4.3 M_\odot (K\ km\ s^{-1}\ pc^2)^{-1}$, but for ULIRGs we also show the effect of assuming $\rm \alpha = 1 M_\odot (K\ km\ s^{-1}\ pc^2)^{-1}$.

Concerning the gas metalicity in the local spiral sample, there are only a few objects (six after removing possible AGN-dominated galaxies) out of the SINGS sample with submm data  that also have integrated spectroscopy \citep{moustakas07,kennicutt92}, which could be used to infer the gas metalicity with the same strong line method and calibration scale adopted for SMGs. For these objects we also exploited the diagnostics involving [OII]3727, [OIII]5007 and H$\beta$, which help to better constrain the metalicity along with
[NII]/H$\alpha$, while they are still fully consistent with the global inter-calibrations presented in \cite{maiolino08}.
\cite{draine07} uses the metalicities for a more extended set of SINGS galaxies, but by adopting different calibrations, which therefore cannot be used for direct comparison with the SMGs.
To expand the sample of objects with dust masses {\it and} metalicity (especially in the low-metalicity range), we included nine dwarf and spiral galaxies  of the SINGS sample without submm
photometry \citep{draine07}. There the $\rm M_{dust}$ estimates were taken from \cite{draine07} (whose dust masses are
consistent with ours for the objects in common, as discussed in the text), and their integrated emission line fluxes, which were required to place them in the same metalicity scale as SMGs and ULIRGs, were collected from \cite{moustakas07}, \cite{miller96} and \cite{miller96b}. 

For the ULIRGs we collected spectroscopic data from \cite{rupke08}. We removed galaxies classified
as AGNs based on the diagnostic diagrams.
 
In order to have a more comprehensive statistic (only 6 out of the 26 local spirals from the SINGS sample have integrated spectroscopy which can
be used to compute their metal abundance),  the metalicity distribution of local spirals shown by the red, solid line in
Fig.~\ref{fig:histo}c is obtained from the large SDSS sample presented in \cite{kewley08}
in the same
mass range as SMGs and ULIRGs,  with the same metalicity calibrations and scale used in \cite{maiolino08}, i.e. consistent with the other samples.
The metalicity distribution of LBGs and BzKs at \textit{z}$\sim$2.2 (with $\rm \log{M_{\star}}\sim 10.6 M_{\odot}$) is taken from \cite{mannucci10},
who adopt the same scale and calibrations as \cite{maiolino08}.

\end{appendix}

\end{document}